\documentclass{appolb}
\usepackage{graphicx}
\usepackage[utf8x]{inputenc}
\usepackage[T1]{fontenc}
\usepackage{hyperref}
\usepackage{amsmath, amssymb, amsfonts} 

\usepackage[style=phys]{biblatex} 
\usepackage{geometry}
\usepackage{algorithm}
\usepackage[noEnd=False]{algpseudocodex}
\usepackage{color}

\usepackage{lipsum}

\begin{document}

\title{On generating Special Quasirandom Structures:\\
Optimization for the DFT computational efficiency}
\author{Andrzej P. Kądzielawa
\thanks{\href{mailto:andrzej.kadzielawa@agh.edu.pl}{andrzej.kadzielawa@agh.edu.pl}}
\address{AGH University of Krakow, Faculty of Physics and Applied Computer Science,\\
al. A. Mickiewicza 30, 30-059 Krakow, Poland}
}
\maketitle
\begin{abstract}
	We present our novel evolutionary algorithm for generating Special Quasirandom Structures (SQS) designed to optimize the computational efficiency of Density Functional Theory (DFT) computations. Operating on the premise that symmetry proxies non-randomness, we rigorously filter out 1.P1 candidate structures prior to evaluating correlation functions. Our extinction-based workflow includes the seeding, filtration, evaluation, extinction, and repopulation phases to produce efficient supercells with maximal local environmental distinctness. We compare our results against those generated by established software packages, on the example of the W\textsubscript{70}Cr\textsubscript{30} alloy. Although standard tools achieve (marginally) lower correlation errors, our best-performing structures require approximately five times fewer unique displacements for phonon calculations. This approach sacrifices negligible quantitative disorder accuracy to significantly reduce the computational cost of modeling thermal properties.
\end{abstract}

\section{Introduction}
\label{sec:introduction}

In recent years, there has been a growing incentive for the exploration of compositionally complex and disordered systems. This transition is driven by the urgent demand for materials capable of withstanding extreme environments (e.g., the so-called Lost-of-coolant accident - LOCA), particularly in applications of fusion energy \cite{AbuShawareb2022,Zylstra2022} and nuclear shielding \cite{Saleh,Alzahrani,Tasnim,Wrobel,Souza}. It made the stochastic arrangement of atoms a primary design variable, which is exemplified by the rise of High-Entropy Alloys (HEAs) \cite{Kozelj2014,Jung2022,Leung2022,Sereika2024,Place2021,Sobota2025,Jasiewicz2023,Jasiewicz2024} and refractory-metal alloys \cite{Smith2000,Kurtz2000,Neu2005,Wurster2011,ElAtwani2023,Mo2024,Acemi2024,Wei2024,Hatler2024,Wurmshuber2024,Mackova2023}, where stabilization of the entropy and local chemical tuning are critical. However, the theoretical treatment of such systems presents a major obstacle: a computational complexity overtaking the existent HPC resources, i.e., the modeling of thermal properties (necessary for alloys) and their stability, requires more resources than those available. The current state-of-the-art ab-initio tools are based on Density Functional Theory (DFT) \cite{Hohenberg,Kohn,KohnNobel,Liechtenstein,Dudarev,Kotliar,Vollhardt}, which provides a robust quantum mechanical framework. The reliance on translational symmetry contradicts the intrinsic nature of disordered alloys. The Special Quasirandom Structure (SQS) method \cite{Zunger,Wei,Jiang} resolves this issue by constructing finite periodic supercells that mimic the disorder of an infinite random alloy. Given that the computational cost of DFT scales cubically with the size of the system ($O(N^3)$), the central optimization problem is not only minimizing the correlation error in the infinite limit, but identifying the set of most computationally efficient optimized supercells that capture the essential physics.

In the latest developments, new SQS algorithms have been made available, together with their implementations, e.g., \texttt{SimplySQS} by Lebeda et al. (2025) \cite{Lebeda2025} or \texttt{PyHEA} by Niu and Liu (2025) \cite{Niu2025}. Nevertheless, these approaches focus more on optimization of the workflow (via massive parallelization using GPU \cite{Niu2025} with the impressive 250,000-atom supercells) or user experience (through a web-based framework \cite{Lebeda2025}). Although their efficiency overshadows the state-of-the-art \texttt{ATAT} \cite{ATAT} and \texttt{sqsgenerator} \cite{sqsgen}, they do not change the underlying algorithm. In addition, frameworks such as \texttt{AlloyGAN} \cite{Hao2025} and material network representations \cite{Zhang2025} employ deep learning to instantaneously construct valid alloy architectures, shifting the bottleneck entirely to the physical solver. 

Previously, we showed that the focus on computational complexity when generating SQSs allows incorporation of thermal effects that reproduce both miscibility gaps in W-Cr-Mo binary systems \cite{Kadzielawa2023} and predict the stability and elastic properties of tungsten-based refractory-metal alloys (W-Cr-Ta/Hf) \cite{Veverka2}.

In this paper, we present our evolutionary approach that operates on the premise that symmetry is a proxy for non-randomness. By rigorously filtering out high-symmetry and redundant candidates before the computationally expensive evaluation of correlation functions, the algorithm accelerates the discovery of optimal structures. Crucially for DFT efficiency, this method produces small supercells with maximal local environmental distinctness. This trade-off, even though reducing the randomness globally, does not impact the local disorder "as seen" from a random lattice node. Moreover, this algorithm allows for corse-grained, massively-parallel, stochastic implementation.

In Section~\ref{sec:introduction} we describe the theoretical foundations of our approach, as well as the underlying algorithm. Then, in Sec.~\ref{sec:analysis} we compare our resultant structures with those obtained from established software \cite{ATAT,sqsgen}, using the alloy W\textsubscript{70}Cr\textsubscript{30} as an example. Finally, in Sec.~\ref{sec:conclusions} we summarize the findings presented in this paper.

\section{Theoretical Foundations of Configurational Disorder}
\label{sec:theoretical}

In this chapter, we introduce a population-based evolutionary approach driven by an \emph{extinction} mechanism.

\subsection{The Cluster Expansion Formalism}
\label{ssec:ce}
The basis of the SQS method is the \emph{Cluster Expansion (CE)} formalism \cite{ClusterExpansion}. For a binary alloy $A_{1-x}B_x$, we assign a variable $\sigma_i$ to each site $i$ (e.g., $0$ for the atom $A$, $1$ for the atom $B$). Any physical property $P$ of a configuration $\vec{\sigma}$ can be expanded
\begin{equation}
	P(\vec{\sigma}) = P_0 + \sum_{\alpha} P_{\alpha} \mathfrak{C}_{\alpha}(\vec{\sigma}).
\end{equation}
Here, $\mathfrak{C}_{\alpha}(\vec{\sigma})$ is the structure-dependent correlation function for a cluster $\alpha$. SQS generation is an optimization problem: finding a configuration $\vec{\sigma}$ in a supercell $S$ such that:
\begin{equation}
	\mathfrak{C}_{\alpha}(S) \approx \mathfrak{C}_{\alpha}^{rand}
\end{equation}
for all physically relevant clusters $\alpha$.

\subsection{Extinction-Based Generation Method (Pseudocode)}
\label{ssec:code}
Although standard tools (e.g., \texttt{ATAT}) often (and for large supercells almost exclusively) produce structures with $P1$ (lowest) symmetry, we aim at filtering for residual symmetries (e.g., $C2$, $Amm2$). This decreases the number of irreducible displacements required for phonon calculations without introducing qualitative errors in the disordered limit, which leads to the possibility to apply the so-called quasiharmonic approximation (QHA) \cite{QHA,PHONOPY} to obtain the high-temperature properties of an alloy.

The following Algorithm~\ref{alg:generator} defines \emph{a generic} workflow to generate a set of SQSs using the extinction-based approach. Note that for the algorithm to function, we need to define the lattice type, supercell size, tolerance, and population size. The approach consists of five phases: \emph{(i)} seeding, where the initial cells are generated; \emph{(ii)} filtration, where the low-symmetry cells are filtered out; \emph{(iii)} evaluation, where correlation/error functions are calculated; \emph{(iv)} extinction, where the worst performing cells are rejected; and \emph{(v)} repopulation, where a new set of offspring cells are generated from the survivors.

\begin{algorithm}
	\caption{Extinction-Based SQS Generator. This is the simplest possible algorithm. In general, we allow for a small $\ll P$ number of the best $P1$ structures modifying line \ref{alg:line:P1}. We also include a standard Metropolis algorithm \cite{Metropolis} (with the reciprocal \emph{temperature} $\beta$), to allow for some locally \emph{nonoptimal} states, modifying \ref{alg:line:prob}.}
	\begin{algorithmic}[1]
		\Require Lattice type $L$, Supercell size $N$, Tolerance $\mathfrak{E}_{tol}$, Population size $P$
		\Statex \textbf{Input:} Concentrations $\{x_i\}$, Output size $M$
		\State $Population \gets \emptyset$
		\While{\Call{Length}{$Population$} $< M$} \Comment{Phase i: Seeding}
		\State $C \gets$ \Call{GenerateRandomSupercell}{$L, N, \{x_i\}$}
		\If{\Call{Symmetry}{$C$} $\neq P1$ \textbf{and} $C$ is unique} \Comment{Phase ii: Filtration} \label{alg:line:P1}
		\State \Call{Population.append}{$C$}
		\EndIf
		\EndWhile
		\State $\mathfrak{E}_{min} \gets \infty$
		\While{$\mathfrak{E}_{min} > \mathfrak{E}_{tol}$} \Comment{Phase iii-v: Evolution}
		\For{$S_i$ in $Population$} \Comment{Phase iii: Evaluation}
		\State $S_i.\mathfrak{E}\gets$ \Call{CalcCorrelationMismatch}{$S_i$} \Comment{ $\mathfrak{E}$ includes displacements.} \label{alg:line:correlation}
		\State $\mathfrak{E}_{min} \gets $ \Call{min}{$\mathfrak{E}_{min}, S_i.\mathfrak{E}$}
		\State $D_{min} \gets $ \Call{num\_of\_unique\_displacements}{$S_i$}
		\EndFor
		\If{$\mathfrak{E}_{min}/D_{min} \leq \mathfrak{E}_{tol}$}
		\Return $Population$ \label{alg:line:break}
		\EndIf
		\State $Survivors \gets \emptyset$
		\For{$S_i$ in $Population$} \Comment{Phase iv: Extinction}
		\State $P_{survival} \gets \mathfrak{E}_{min} / S_i.\mathfrak{E}$ \Comment{Can be replaced with $\exp(-\beta( S_i.\mathfrak{E} - \mathfrak{E}_{min} )$} \label{alg:line:prob}
		\If{\Call{Random}{0,1} $\leq P_{survival}$}
		\State \Call{Survivors.append}{$S_i$}
		\EndIf
		\EndFor
		\State $Population \gets Survivors$
		\While{$\text{Length}(Population) < P$} \Comment{Phase v: Repopulation}
		\State $\text{Parent} \gets $ \Call{SelectRandom}{Survivors}
		\State $\text{Offspring} \gets $ \Call{Mutate}{Parent}
		\If{\Call{Symmetry}{Offspring}$ \neq P1$}
		\State \Call{Population.append}{Offspring}
		\EndIf
		\EndWhile
		\EndWhile
		\Statex \textbf{Output:} $M$ Optimal SQS Structures
	\end{algorithmic}
	\label{alg:generator}
\end{algorithm}

The separate calculation in Alg.~\ref{alg:generator}:\ref{alg:line:correlation} is described in more detail as Algorithm~\ref{alg:correlation}. Note that the required weights $A_n$ are a series of (typically) decreasing monotonically with $n$ values that prioritize correlations for the nearest-neighbors. As there is no \emph{ab-initio} form of $A_n$, we choose the power form of $A_n \equiv \tfrac{1}{n}$. Tests with $A_n \equiv n^{-p}$, with $p \in \{2,4,12\}$ and $A_n = R_n^{-1}$ ($R_n$ being the distance to the coordination zone) produced similar results for the P-1 structures, but performed worse for computationally efficient supecells (with a convergence of the algorithm being of an order of magnitude slower).

\begin{algorithm}
	\caption{Calculating error: \texttt{CalcCorrelationMismatch}}
	\begin{algorithmic}[1]
		\Require Weights $A_n$
		\Statex \textbf{Input:} Supercell, Concentrations $x$, 
		\State $BackgroundSupercell \gets  \emptyset$
		\State $a,b,c = Supercell.directions$
		\For{i,j,k in \Call{Carthesian\_product}{$\{-1,0,1\}$, 3}}
		\For{atom in Supercell}
		\State $mirrorAtom \gets atom$
		\State $mirrorAtom.position \gets atom.position + i\,a + j\,b + k\,c$
		\State $BackgroundSupercell.append(mirrorAtom)$
		\EndFor
		\EndFor
		\State $CoordZones \gets$ \Call{Find\_coordination\_zones}{Supercell,BackgroundSupercell}
		\For{Zone in CoordZones}
		\For{atomType in Zone.atomTypes}
		\State $Zone[atomType].correlation \gets 0$
		\EndFor
		\EndFor
		\For{atomS in Supercell}
		\For{atomB in BackgroundSupercell}
		\If{atomS.type == atomB.type}
		\State \Call{CoordZones.from\_atoms}{atomS,atomB}[atomB.type].correlation $\texttt{+=1}$
		\EndIf
		\EndFor
		\EndFor
		\State $\mathfrak{E} \gets 0$
		\For{n,Zone in \Call{enumerate}{CoordZones}}
		\For{atomType in Zone.atomTypes}
		\State $Zone[atomType].correlation \texttt{/=} Zone[atomType].population$
		\State $Zone[atomType].correlation \texttt{-=} x[atomType]$
		\State $\mathfrak{E} \texttt{+=} A_n \times |Zone[atomType].correlation|$ \Comment{No $||\rightleftharpoons$ error cancellation included}
		\EndFor
		\EndFor
		\State $\mathfrak{E} \texttt{*=} $ \Call{num\_of\_unique\_displacements}{Supercell}\Comment{Optimization; This favors computationally "cheaper" supercells.} \label{alg:line:disp}
		\Statex \textbf{Output:} Mismatch magnitude $\mathfrak{E}$
	\end{algorithmic}
	\label{alg:correlation}
\end{algorithm}

Note that in Alg.~\ref{alg:correlation}:\ref{alg:line:disp} we modify the error function to penalize structures with the large number of unique displacements required for the phononic spectrum computations. This ensures that Alg.~\ref{alg:generator} uses the (computationally) cheaper structures, cf. the success condition in Alg.~\ref{alg:generator}:\ref{alg:line:break} does not take the number of displacements into account. In the next section, we compare our results with those obtained with state-of-the-art tools.

\section{Comparative Analysis}
\label{sec:analysis}
In this section, we analyze the error function of our resultant supercells compared to those generated using the established software.

\begin{figure*}
	\centering
	\includegraphics[width=\linewidth]{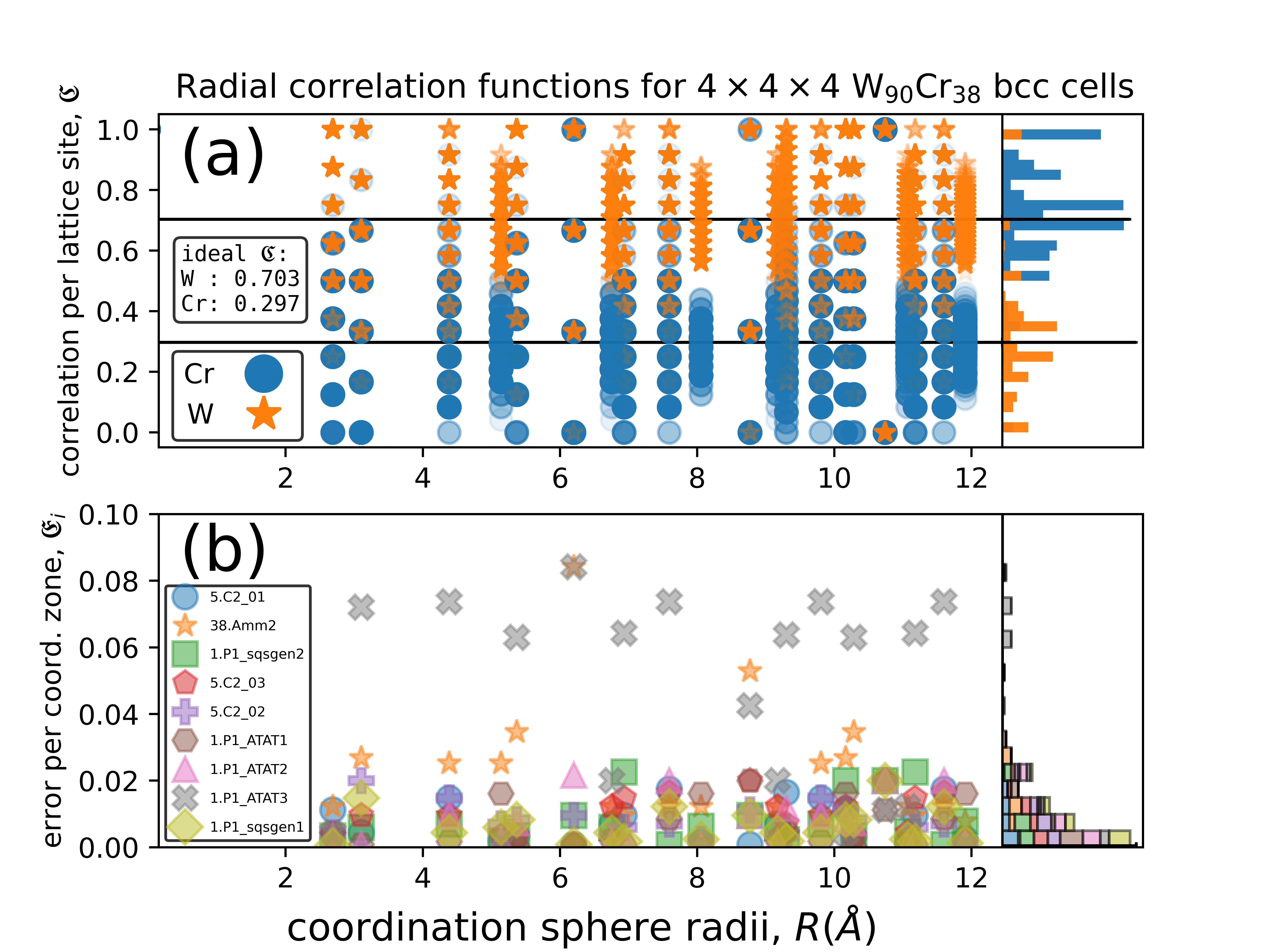}
	\caption{Correlation \textbf{(a)} and error \textbf{(b)} functions for different $4\times 4\times 4$ bcc Special Quasirandom W\textsubscript{70}Cr\textsubscript{30} Structures (exactly W\textsubscript{90}Cr\textsubscript{38} supercell). Histograms on the right sum up the corresponding populations. We used two \texttt{sqsgenerator} \cite{sqsgen} generated 1.P1-symmetry cells with 10\textsuperscript{9} (1.P1\_sqsgen1) and 10\textsuperscript{9} (1.P1\_sqsgen2) iterations, and three \texttt{ATAT} \cite{ATAT} 10\textsuperscript{9} (1.P1\_ATAT1 and 1.P1\_ATAT2), and 10\textsuperscript{2} (1.P1\_ATAT3) iterations. We included our own cells from 5.C2 and 38.Amm2 symmetry groups, while the result for purely random 1.P1 cell is given as a reference point. Note that, while 1.P1\_ATAT1 (brown hexagons) is the best (error-wise with $\mathfrak{E} = 2.01 \,\cdot 10^{-4}$), our 5.C3\_03 ($\mathfrak{E} = 2.92 \,\cdot 10^{-4}$, red pentagons) has $\sim 5$ times fewer displacements for phononic calculations.}
	\label{fig:correlation}
\end{figure*}

The key feature of the approach above is to \emph{(i)} \textbf{get a set of computationally effective supercells} that still \emph{(ii)} \textbf{model the local disorder} of an alloy (still being SQSs). While \emph{(i)} is satisfied directly, to substantiate \emph{(ii)} we verify the "properness" of our SQSs by comparing them to those generated by established codes (here \texttt{sqsgenerator} \cite{sqsgen} and \texttt{ATAT} \cite{ATAT}) in terms of the pure error function (without the inclusion of the number of displacement). As pictured in Fig.~\ref{fig:correlation}, while the best result is achieved by the \texttt{ATAT} suite (olive diamonds), our best performer (green squares) gets similar results for the first five coordination zones (the latter deviate from optimal disorder).

\section{Conclusions}
\label{sec:conclusions}
We have proposed a modification of an established state-of-the-art tool for describing the \emph{ab-initio} properties of alloys. By sacrificing some quantitative efficiency of describing lattice disorder, we obtain a set of Special Quasirandom Structures that perform well computationally while still modeling alloy's disorder. As in predicting new stable alloys, as well as assessing the decomposition rates \cite{Veverka2}, we need not only to calculate optimal structures for the target composition, but also its neighborhood in the composition--enthalpy-of-formation vector space. This is necessary to predict stability (as a function of external pressure $p$ and temperature $T$) by studying the convex hull (CH) in the aforementioned vector space and whether the composition-of-interest lies in the convex hull (stability) or not (the distance from CH determines the decomposition rate).

\section{Acknowledgments}
The author acknowledges the continued support of Prof. Józef Spałek (Institute of Theoretical Physics, Jagiellonian University), whose insightful discussions helped to develop this research. As this work coincides with the 50-year scientific career jubilee, we wish to pay special tribute to his enduring mentorship and scientific vision. We gratefully acknowledge Poland's high-performance Infrastructure PLGrid ACK Cyfronet AGH for providing computer facilities and support within computational grant no PLG/2025/018497. Finally, the author recognizes the help of Dominik Legut and Sergiu Arapan (IT4Innovations, V\v{S}B - Technical University of Ostrava) in conceptualizing the problem described in this paper.

\printbibliography

\newpage
\end{document}